\documentclass[prl,twocolumn,superscriptaddress,showpacs]{revtex4}
\usepackage{amsmath}
\usepackage{amsfonts}
\usepackage{graphicx}

\begin{document}

\title{Geometric Phase of Three--level Systems in Interferometry}
\author{Barry C.\ Sanders}
\affiliation{Department of Physics, Macquarie University, Sydney,
        New South Wales 2109, Australia}
\author{Hubert \surname{de Guise}}
\affiliation{Department of Physics, Macquarie University, Sydney,
        New South Wales 2109, Australia}
\affiliation{Centre de Recherche Math\'ematique, Universit\'e de
        Montr\'eal, C.P.\ 6128--A,
        Montr\'eal, Qu\'ebec H3C 3J7, Canada}
\author{Stephen D.\ Bartlett}
\affiliation{Department of Physics, Macquarie University, Sydney,
        New South Wales 2109, Australia}
\author{Weiping Zhang}
\affiliation{Department of Physics, Macquarie University, Sydney,
        New South Wales 2109, Australia}

\date{November 14, 2000}

\begin{abstract}
  We present the first scheme for producing and measuring an Abelian
  geometric phase shift in a three--level system where states are
  invariant under a non-Abelian group.  In contrast to existing
  experiments and proposals for experiments, based on U(1)--invariant
  states, our scheme geodesically evolves U(2)--invariant states in a
  four--dimensional SU(3)/U(2) space and is physically realized via a
  three--channel optical interferometer.
\end{abstract}

\pacs{03.65.Bz, 02.20.-a, 42.50.-p}
\maketitle

Cyclic evolution of a wave function yields the original state plus a
phase shift, and this phase shift is a sum of a dynamical phase and a
geometric (or topological, or quantal, or Berry) phase
shift~\cite{Ber84,Sim83}.  The geometric phase shift is important, not
just for quantum systems, but also for all of wave physics.  Thus far,
controlled geometric--phase experiments, both realized and proposed,
have been exclusively concerned with the so--called Abelian geometric
phase arising in the evolution of U(1)--invariant states, for example,
states of the Poincar\'{e} sphere (in the case of SU(2)
states)~\cite{Tom86,Kwi91}, the Lobachevsky plane~\cite{Sim93} (in the
case of SU(1,1) states) and ${\Bbb R}^2$ (for the Aharonov--Bohm
phase)~\cite{Ber84}.  Here we introduce an optical scheme to produce
and detect an Abelian geometric phase shift which arises from geodesic
transformations of U(2)--invariant states in a four--dimensional
SU(3)/U(2) space.  This scheme employs a three--channel optical
interferometer and four experimentally adjustable parameters to
observe the geometric phase in its full generality.

Geometric phases in SU(3) systems have been the subjects of recent
mathematical studies~\cite{Muk93,Arv97} and establish the geometric
phase shift expected for the cyclic evolution (up to a phase) of a
three--level system.  We propose to obtain this evolution using an
interferometer as a sequence of unitary transformations given by
optical elements.  An optical SU(3) transformation can be realized by
a three--channel optical interferometer~\cite{Rec94}.  The space of
output states of the interferometer can be identified with SU(3)/U(2),
and will be referred to as the geometric space; this space is a
generalization of the Poincar\'e sphere to a three--level
system~\cite{Arv97}.  By adjusting the parameters of the
interferometer, the output state can be made to evolve cyclically, up
to a phase, through a triangle in the geometric space.  The output of
the interferometer may be any state along a path in SU(3)/U(2),
determined by fixing the four free parameters of the interferometer.

It is important to distinguish the evolution of states in the
geometric space SU(3)/U(2) from the transformations of the optical
beam as it progresses through the interferometer.  It will be shown
later how the dynamical phase associated with these optical
transformations can be eliminated.  The cyclic evolution described in
this paper occurs in the geometric space, and the geometric phase of
interest is related to this evolution.  We provide here the essential
elements to obtain this evolution as well as to explain how to design
the interferometer.

It is sufficiently general to consider the input state $\psi^{\rm in}$
of a photon into one of the three input ports and the vacuum state
into the other two ports.  The parameters of the interferometer can be
initially set such that the resultant SU(3) transformation is the
identity, and thus the interferometer output state $\psi^{(1)}$ is
also one photon at the corresponding output port and the vacuum at the
two other ports.  These parameters can then be adjusted to evolve the
output state along a trajectory in the geometric space; this evolution
may involve both a `dynamical' phase shift and a geometric phase
shift.  Care must be taken when interpreting the adjective
`dynamical'.  The output state does not evolve according to
Schr\"odinger dynamics but instead follows a path in the geometric
space parametrized by an evolution parameter $s$, which is a function
of the adjustable parameters of the interferometer.

\begin{figure}
  \includegraphics*[width=3.25in,keepaspectratio]{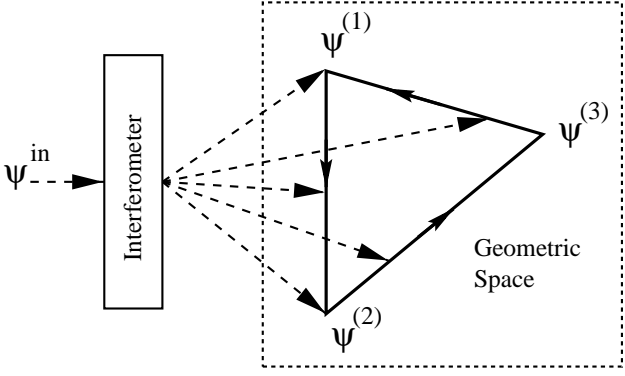}
        \caption{
          The geodesic evolution in the geometric space is depicted
          diagrammatically.  By adjusting the parameters of the
          interferometer, the output state in the geometric space can
          be made to evolve along geodesic paths, from one vertex to
          the next, until the triangle is closed.  The SU(3)
          transformations $U^g_k(s_k)$ map the output state along the
          geodesic paths in the geometric space.  }
  \label{fig:geodesic}
\end{figure}

The dynamical and geometric phase shift contributions must be
separated to obtain the geometric phase.  A special role is played by
geodesic evolution~\cite{Chy88}; by transforming the output state
along geodesic paths in the geometric space, the geometric phase shift
along each path is zero.  Thus, we consider three arbitrary states $\{
\psi^{(k)}; k=1,2,3 \}$ in the geometric space which define a geodesic
triangle (i.e., with sides given by the unique geodesics connecting
these states).  The parameters of the interferometer are adjusted to
evolve the output state along this general geodesic triangle
$\psi^{(1)} \to \psi^{(2)} \to \psi^{(3)} \to \psi^{(4)} =
e^{i\varphi_g}\psi^{(1)}$, where $\varphi_g$ is the total geometric
phase gained by cyclic evolution and depends on four free parameters
of the interferometer.  Fig.~\ref{fig:geodesic} gives a diagrammatic
depiction of this scheme.

The evolution of the state~$\psi^{(1)}$ to the state $\psi^{(4)} =
e^{i\varphi_g}\psi^{(1)}$ via three geodesic paths in the geometric
space can be described by three one--parameter SU(3) group elements
$\{ U^g_k(s_k);\ k=1,2,3 \}$, with $s_k$ an evolution parameter.
These transformations satisfy the conditions that $U^g_k(0)$ is the
identity element and
\begin{equation} 
  \label{U_k}
  U^g_k(s_k^0) \psi^{(k)} = \psi^{(k+1)} , \quad k =  1,2,3 \, ,
\end{equation}
for some fixed values $\{ s_k^0 \}$.  It is always possible to choose
unit vectors $\psi^{(k)}$ such that $\langle \psi^{(k+1)}\vert
\psi^{(k)}\rangle$ is real and positive.  We consider evolutions
$U_k^g(s_k)$ of the form
\begin{equation}
  \label{eq:DecomposeU}
  U_k^g(s_k)= V_k\cdot R_{s_k}\cdot V_k^{-1}\, ,
\end{equation}
with $V_k$ an element of SU(3) satisfying $\langle \psi^{(k)} | U^g_k(s_k)
| \psi^{(k)} \rangle$ real and positive, and
\begin{equation} 
  \label{eq:rt}
  R_{s_k}       \equiv \left( \begin{array}{ccc}
                \cos s_k & -\sin s_k & 0 \\
                \sin s_k & \cos s_k & 0 \\
                0&0&1\end{array}\right)\, .
\end{equation}
The form of the one--parameter subgroup $R_{s_k}$ with real entries
was guided by the definition of a geodesic curve between two states
$\psi^{(k)}$ and $\psi^{(k+1)}$, which can be written in the
form~\cite{Arv97}
\begin{eqnarray}
  \label{eq:geocurve}
  \psi(s_k) &=& \psi^{(k)}\cos s_k \\
  &&+{ \left(\psi^{(k+1)}-\psi^{(k)}
  \langle \psi^{(k+1)}\vert \psi^{(k)}\rangle \right) \over 
  \sqrt{1- \langle \psi^{(k+1)}\vert \psi^{(k)}\rangle^2 }}
  \sin s_k \nonumber
\end{eqnarray}
with $0\le s_k \le s_k^0 = \arccos \langle \psi^{(k+1)}\vert
\psi^{(k)}\rangle$.  It is straightforward to show that any
$U^g_k(s_k)$ of the form given by Eq.~(\ref{eq:DecomposeU})
satisfying $\langle \psi^{(k+1)}\vert \psi^{(k)}\rangle$ real and
positive gives evolution along a geodesic curve in SU(3)/U(2).

Consider the three states
\begin{eqnarray}
  \label{vertices}
  \psi^{(1)} &=& \left( \begin{array}{c} 1 \\ 0 \\ 0 \end{array} \right)
        = e^{-i\varphi_g} \psi^{(4)} , \quad
  \psi^{(2)} = \left( \begin{array}{c} \cos s_1^0 \\ \sin s_1^0 \\ 0 
  \end{array} \right) ,
                \nonumber \\
  \psi^{(3)} &=&  \left( \begin{array}{c}
        \cos s_1^0 \cos s_2^0 - e^{i\alpha} \sin s_1^0 \sin s_2^0
                \cos \beta \\
        \sin s_1^0 \cos s_2^0 + e^{i\alpha} \cos s_1^0 \sin s_2^0
                \cos \beta \\
        \sin \beta \sin s_2^0
        \end{array} \right) ,
\end{eqnarray}
with $s_1^0$, $s_2^0$, $\alpha$ and $\beta$ arbitrary.  These three
states form the vertices of the geodesic triangle in the geometric
space. They are sufficiently general to include all types of geodesic
triangles~\cite{Arv97}.

Although the three--channel interferometer can be expressed as an
SU(3) transformation (or sequence of SU(3) transformations), the
optical elements of the interferometer are composed of beam splitters,
mirrors and phase shifters.  Provided that losses can be ignored, each
of these optical elements can be associated with an SU(2) unitary
transformation~\cite{Yur86,Cam89}.  It is therefore advantageous to
factorize each SU(3) transformation into a product of SU(2)$_{ij}$
subgroup transformations mixing fields $i$ and $j$: first, an
SU(2)$_{23}$, followed by an SU(2)$_{12}$ and completed by a final
SU(2)$_{23}$ transformation~\cite{Row99}.  Such a factorization makes
the experimental design of the interferometer clear: fields $2$ and
$3$ are mixed followed by a mixing of the output field~$2$ with the
field in channel~$1$, and, finally, the output field~$2$ is mixed with
field~$3$.

The SU(2)$_{12}$ matrix~$R_s$ in Eq.~(\ref{eq:rt}) is a special case
of the generalized lossless beam splitter transformation for mixing
channels~$1$ and~$2$.  More generally a beam splitter can be described
by a unitary transformation between two channels~\cite{Cam89}.  For
example, a general SU(2)$_{23}$ beam splitter transformation for
mixing channels~$2$ and $3$ is of the form
\begin{equation}
  R_{23}(\phi_{\rm t},\theta,\phi_{\rm r})
        = \left(\begin{array}{ccc}
        1&0&0\\
        0&e^{i\phi_{\rm t}}\cos\theta&-e^{-i\phi_{\rm r}}\sin\theta\\
        0&e^{i\phi_{\rm r}}\sin\theta
                &e^{-i\phi_{\rm t}}\cos\theta\end{array}
        \right) \, ,
\end{equation}
with~$\phi_{\rm t}$ and $\phi_{\rm r}$ the transmitted and reflected
phase--shift parameters, respectively, and $\cos^2\theta$ the beam
splitter transmission.  A generalized beam splitter can be realized as
a combination of phase shifters and 50/50 beam splitters in a
Mach--Zehnder interferometer configuration.

It is useful at this point to consider the nature of the geodesic
transformations $U^g_k(s_k)$ and their realization in terms of optical
elements.  The interferometer can be adjusted to transform the input
state $\psi^{\rm in}$ to an arbitrary output state $\psi(s)$ anywhere
along the geodesic triangle.  This optical transformation can be
related to an SU(3) transformation in the geometric space, which maps
$\psi^{(1)}$ to $\psi(s)$ along a geodesic path.  It is important to
distinguish between the optical evolution through the interferometer
from $\psi^{\rm in}$ to $\psi(s)$, and the geodesic evolution in the
geometric space from $\psi^{(1)}$ to $\psi(s)$.  The goal of the
following is to construct optical transformations in terms of SU(2)
elements which realize the geodesic evolution in the geometric space
by appropriately adjusting parameters.

It will be convenient to express $\psi^{(3)}$ as $(e^{i\xi}\cos\eta,
e^{i(\xi+\chi)} \sin\eta\cos\tau, \sin\eta\sin\tau)^T$, where $\xi$,
$\eta$, $\tau$ and $\chi$ are functions of $s_1^0$, $s_2^0$, $\alpha$
and $\beta$, the parameters of $\psi^{(3)}$ in Eq.~(\ref{vertices}).
Following our factorization scheme, the geodesic evolution operators
$U^g_k(s_k)$, connecting $\psi^{(k)}$ to $\psi^{(k+1)}$, can be
expressed as
\begin{eqnarray}
  \label{evolutions}
  U^g_1(s_1)
        &=& R_{s_1}\, ,\nonumber \\
  U^g_2(s_2)
        &=& R_{s_1^0}\cdot R_{23}(\alpha,\beta,0)\cdot R_{s_2}  
        \cdot R^{-1}_{23}(\alpha,\beta,0)\cdot R_{-s_1^0}\, 
        ,\nonumber \\
  U^g_3(s_3)&=&
        R_{23}(\chi,\tau,-\xi) \cdot R_{-s_3} \cdot 
        R^{-1}_{23}(\chi,\tau,-\xi)
        \, ,
\end{eqnarray}
with $R_s$ given by Eq.~(\ref{eq:rt}), the parameters $s_k$ ranging
from $0\leq s_k \leq s_k^0$, and $s_3^0 = \eta$.  Note that $s_3^0$
and, in fact, all the parameters of $U^g_3(s_3)$ are fixed by the
requirement that $\psi^{(4)} = e^{i\varphi_g}\psi^{(1)}.$ Also note
that, for each $k$, $U^g_k(0)$ is the identity in SU(3) and
$U^g_k(s_k^0)\psi^{(k)} = \psi^{(k+1)}$ as required.  Once it is
observed that $\langle \, \psi^{(k+1)}\vert \ \psi^{(k)}\, \rangle =
\cos s_k^0$, it is trivial to verify that each evolution satisfies
Eq.~(\ref{eq:geocurve}) and is therefore geodesic.

The geometric phase for the cyclic evolution $\psi^{(1)} \to 
\psi^{(4)}$ is given explicitly by
\begin{equation}
  \label{geometricphase}
  \varphi_g = \xi = {\rm arg}(\cos s_1^0 \cos s_2^0 - 
  e^{i\alpha}\sin s_1^0 \sin s_2^0 \cos \beta) \, .
\end{equation}
This phase depends on four free parameters in the experimental scheme:
$s_1^0$, $s_2^0$, $\alpha$ and $\beta$, which describe a general
geodesic triangle in SU(3)/U(2).

\begin{figure}
  \includegraphics*[width=3.25in,keepaspectratio]{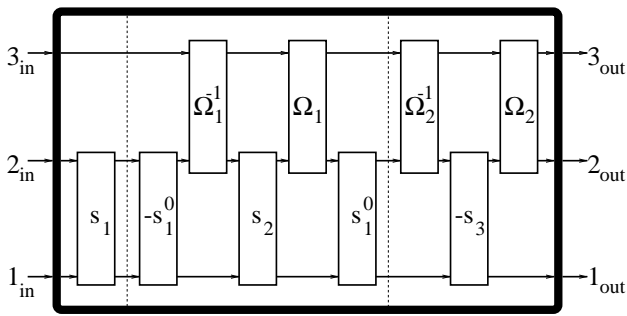}
        \caption{
          The SU(3) interferometer is depicted, with three input ports
          $1_{\rm in}$, $2_{\rm in}$ and $3_{\rm in}$, and three
          output ports $1_{\rm out}$, $2_{\rm out}$ and $3_{\rm out}$.
          There are nine beam splitter transformations with parameters
          $s_1$, $s_2$, $s_3$ and $\Omega_1 = (\alpha,\beta,0)$,
          $\Omega_2 = (\chi,\tau,-\xi)$.  For geodesic, cyclic
          evolution of the output state, only four parameters are
          independent.  }
  \label{fig:config}
\end{figure}

The interferometer configuration for realizing the necessary evolution
about the geodesic triangle is depicted in Fig.~\ref{fig:config}.
This configuration consists of a sequence of SU(2)$_{ij}$
transformations, which are physically realized by generalized beam
splitters (e.g., Mach--Zehnder interferometers).  We use the shorthand
notation $\Omega_i \equiv (\alpha_i , \beta_i , \gamma_i )$ to
designate the three parameters associated with a generalized beam
splitter.  The three--channel interferometer consists of a sequence of
nine SU(2)$_{ij}$ transformations.  The field enters port~$1_{\rm
  in}$, and the vacuum state enters ports~$2_{\rm in}$ and~$3_{\rm
  in}$.  By adjusting the parameters of the interferometer, the output
state can be made to evolve along the geodesic triangle $\psi^{(1)}
\to \psi^{(4)}$ in the geometric space.  We now consider how to
measure the geometric phase as a function of the four free parameters
$s_1^0$, $s_2^0$, $\alpha$ and $\beta$ describing a general geodesic
triangle.

One check on the proper functioning of the interferometer is to place
photodetectors at the ports $1_{\rm out}$, $2_{\rm out}$, and $3_{\rm
  out}$.  The measured photodistribution for any output state can be
compared to the predicted output $\psi(s)$ of the interferometer.  In
particular, for cyclic evolution to the output state $\psi^{(4)}$,
there should be no photons exiting ports $2_{\rm out}$ and~$3_{\rm
  out}$ regardless of the settings of the free parameters.

Consider the cyclic evolution of the output state to the state
$\psi^{(4)} = e^{i\varphi_g}\psi^{(1)}$.  A key technical challenge is
measuring $\varphi_g$, because one must have a reference state with
which to interfere the output state $\psi^{(4)}$.  The input state
$\psi^{\rm in}$ is a poor choice, because the relationship between
$\psi^{\rm in}$ and $\psi^{(1)}$ involves an optical dynamical phase
due to evolution through the interferometer.  However, this optical
phase can be eliminated through the use of a counter--progagating
beam, described below.

\begin{figure}
  \includegraphics*[width=3.25in,keepaspectratio]{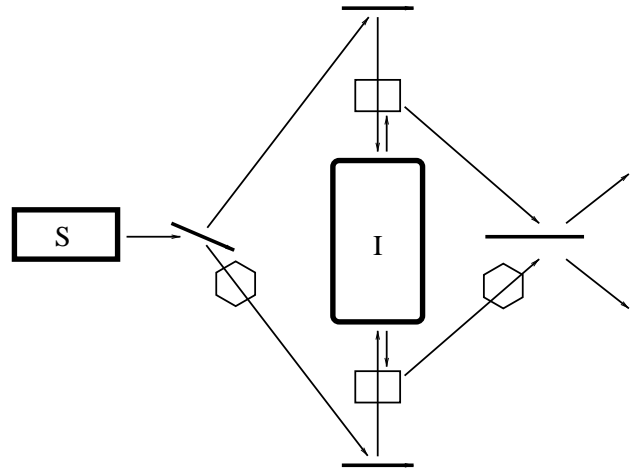}
        \caption{
          In the interferometric scheme for detecting the geometric
          phase shift, the source (S) field is split by a 50/50 beam
          splitter into two identically polarized output fields with
          equal amplitudes.  The polarization of one output field is
          rotated to an orthogonal polarization, with the polarizer
          rotator represented by a hexagon.  One field is injected
          into port~$1_{\rm in}$ of the SU(3) interferometer in
          Fig.~\ref{fig:config}, and the orthogonally--polarized field
          is injected into port~$1_{\rm out}$ at the other end of the
          interferometer.  The output fields exit ports~$1_{\rm out}$
          and~$1_{\rm in}$, respectively and are separated from the
          input fields by polarizing beam splitters at each of the two
          output ports, followed by mixing at a 50/50 beam splitter.
          }
  \label{fig:sagnac}
\end{figure}

A scheme for conducting such an experiment is depicted in
Fig.~\ref{fig:sagnac}.
The source (a laser, for example) produces a
polarized, stable, coherent beam of light which is split at a
polarization--independent beam splitter.  One beam travels to input
port~$1_{\rm in}$ as shown in Fig.~\ref{fig:config} and passes
through the interferometer, exiting at output port~$1_{\rm out}$.  The
other beam is first ``rotated'' to an orthogonal polarization; it then
enters port~$1_{\rm out}$ and counter--propagates through the
interferometer, exiting at port~$1_{\rm in}$.  The orthogonal
polarizations of the two counter--propagating beams ensure that they
do not interfere with each other inside the interferometer.

At the ports~$1_{\rm in}$ and~$1_{\rm out}$, there are polarizing beam
splitters which deflect the outcoming beams but do not affect the
propagation of the incoming beams.  The output beams are directed to a
beam splitter where they are made to interfere.  The optical dynamical
phase shift accumulated by each of the two counter--propagating beams
through the SU(3) interferometer is identical, because of the
time--reversal invariance of the Hamiltonian describing the evolution
within the interferometer.  Thus, the optical phase shifts cancel in
the interference.  Unitarity of the interferometer transformation
guarantees that the geometric phase shift is $\varphi_g$ for one beam
and $-\varphi_g$ for the other beam.  Thus, the two beams interfere
with relative phase $2\varphi_g$.

By measuring the geometric phase $\varphi_g$ for various settings of the
free parameters of the interferometer, it is possible to explore the
geometric space with the most general geodesic triangles.  The observed
values can then be compared to the theoretical predictions.

If the source in Fig.~\ref{fig:sagnac} is a laser, operation at a
low--light level can be undertaken to verify that the geometric phase
shift is $\varphi_g$ even if the probability of more than one photon
being present within the system is negligible.  Low--light level
operation, in the regime where the presence of more than one photon in
the entire apparatus at any time is negligible, enables the testing of
the geometric phase shift even when the discreteness of the field
energy cannot be ignored~\cite{Har93}.

A variation of the scheme in Fig.~\ref{fig:sagnac} can also be
considered to verify that the geometric phase occurs for each photon.
Kwiat and Chiao~\cite{Kwi91} conducted a measurement of geometric phase
by employing parameteric down conversion (PDC), with a UV--pumped KDP
crystal, to produce photon pairs.  One photon undergoes a geometric
phase shift, and the second photon in the pair is employed as a gate
to register the event.  By repeating this process for many `single'
photons, conditioned on detection of the gate photon, where the photon
passes the first beam splitter in Fig.~\ref{fig:sagnac} and has an
equal probability of propagating or, in an orthogonally polarized
state, counter--propagating, through the three-channel interferometer,
an interference pattern can be built up one photon at a time to
establish that geometric phase is imposed one photon at a time,
following Dirac's dictum that ``each photon interferes only with
itself''~\cite{Dir58}.

Although SU(3) interferometry has been considered in detail, the
methods employed here can be extended to SU(N), or N--channel,
interferometry~\cite{Rec94}.  The schemes discussed above employing
such a device would produce and enable observation of the geometric
phase shift for geodesic transformations of states invariant under
\mbox{U(N-1)} subgroups of \mbox{SU(N)} states in the
\mbox{2(N-1)}--dimensional coset space \mbox{SU(N)/U(N-1)}.

\begin{acknowledgments}
  This work has been supported by two Macquarie University Research
  Grants and by an Australian Research Council Large Grant.  BCS
  appreciates valuable discussions with J.\ M. Dawes and A.\ 
  Zeilinger, and HdG acknowledges the support of Fonds F.C.A.R. of the
  Qu\'ebec Government.
\end{acknowledgments}

\end{document}